\documentclass[%
 reprint,
 amsmath,amssymb,
 aps,
]{revtex4-1}

\usepackage{graphicx}
\usepackage{dcolumn}
\usepackage{bm}
\usepackage{array}
\usepackage{xcolor}
\usepackage{svg} 
\usepackage{amsmath} 
\usepackage[normalem]{ulem}
\usepackage{txfonts}
\usepackage{tabularx}
\usepackage{caption}
\usepackage{pifont}

\newcommand{\rev}[1]{\textcolor{black}{#1}} 

\makeatother

\begin{document}

\preprint{APS/123-QED}

\title{\rev{Influence of erythrocyte density on aggregability as a marker of cell age}: Dissociation dynamics in extensional flow}



\author{Midhun {Puthumana Melepattu}}
\email{Midhun.Puthumana-Melepattu@univ-grenoble-alpes.fr}
 \author{Guillaume Ma\^{\i}trejean} 
 \email{Guillaume.Maitrejean@univ-grenoble-alpes.fr}
\author{Christian Wagner}
\email{christian.wagner@uni-saarland.de}
\author{Thomas Podgorski}%
 \email{Thomas.Podgorski@univ-grenoble-alpes.fr}
\affiliation{
Université Grenoble Alpes, CNRS, Grenoble INP, LRP, 38000 Grenoble, France}
\altaffiliation[]{The authors are members of LabEx Tec21 (Investissements d'Avenir - grant agreement ANR-11-LABX-0030)}
\affiliation{Universit\"{a}t des Saarlandes, Saarlandes, Germany
}

\date{\today}

\begin{abstract}

Blood rheology and microcirculation are strongly influenced by red blood cell (RBC) aggregation. The aggregability of RBCs can vary significantly due to factors such as their mechanical and membrane surface properties, which are affected by cell aging in vivo. In this study, we investigate RBC aggregability as a function of their density, a marker of cell age and mechanical properties, by separating RBCs from healthy donors into different density fractions using Percoll density gradient centrifugation. We examine the dissociation rates of aggregates in a controlled medium supplemented with Dextran, employing an extensional flow technique based on hyperbolic microfluidic constrictions and image analysis, assisted by a convolutional neural network (CNN). In contrast to other techniques, our microfluidic experimental approach highlights the behavior of RBC aggregates in dynamic flow conditions relevant to microcirculation. Our results demonstrate that aggregate dissociation is strongly correlated with cell density and that aggregates formed from the denser fractions of RBCs are significantly more robust than those from the average cell population. This study provides insight into the effect of RBC aging in vivo on their mechanical properties and aggregability, underscoring the importance of further exploration of RBC aggregation in the context of cellular senescence and its potential implications for hemodynamics. Additionally, it suggests that this technique can complement existing methods for improved evaluation of RBC aggregability in health and disease.

\end{abstract}

\keywords{Blood, red blood cell, aggregation, RBC aggregability, RBC aging, RBC density, Percoll, artificial intelligence, CNN}

\maketitle

\section{Introduction}

Red blood cell (RBC) aggregation is a critical factor influencing blood rheology and circulation \cite{baskurt2011}. Due to the presence of plasma proteins, RBCs tend to reversibly aggregate, forming mostly linear structures known as \emph{rouleaux} favored by their flat biconcave shape. These linear structures can assemble into larger, gel-like networks at low shear rates and are mostly responsible for the strong shear-thinning of blood observed at low shear rates \cite{chien1970}, in conjunction with RBC deformability \cite{chien1967a,lanotte16}. 

Although shear rates in blood circulation generally exceed 100 s$^{-1}$ \cite{Robertson2008}, where aggregation does not impact macroscopic rheology measurements \cite{chien1970}, various studies have shown that aggregation interactions play a crucial role in shaping the structure of RBC suspensions in narrow channels \cite{Bishop2001, Zhang2009}, stabilizing clusters within capillaries \cite{brust14}, and affecting blood perfusion in vascular networks \cite{Reinhart2017}. Indeed, in narrow vessels or channels, small RBC aggregates made up of a few cells can resist to strain rates up to a few 100 s$^{-1}$ \cite{puthumana2024dissociation}.

RBC aggregation is caused by plasma proteins, especially fibrinogen \cite{baskurt2011}, through two proposed mechanisms, both of which being supported by evidence. The first one is a bridging mechanism in which macromolecules such as fibrinogen adsorb (specifically or not) onto RBC membranes and form bridges between adjacent cells \cite{chien1987physicochemical,Chien1973RedCA}. The second mechanism is of purely entropic nature and involves depletion forces \cite{asakura1958} related to the finite size of macromolecules and was quantitatively predicted by Neu and Meiselmann \cite{neu2002}. While variations in plasma composition have an impact on aggregation, e.g. through fibrinogen production in inflammatory conditions \cite{castell1990acute} which is usually reflected in erythrocyte sedimentation rate (ESR) tests \cite{bedell1985}, cellular factors also strongly influence the ability of cells to aggregate (usually referred to as \emph{aggregability} \cite{ baskurt2013erythrocyte}) and the properties of these aggregates.  For instance, previous modelling and experimental studies have shown that cellular parameters such as cell volume or membrane bending and shear moduli have a significant impact on the equilibrium shape of erythrocyte aggregates \cite{flormann2017,hoore2018}. 
RBCs under normal conditions are highly deformable. Under pathological conditions, such as sickle cell anemia and bacterial infections, deformability of RBCs is greatly reduced. Clinical studies have shown that under such conditions aggregability of RBCs is also increased
\cite{tripette2009red,Baskurt1997Red,Chen1996Enhanced,ChongMartinez2003}.  However reduced deformability is not the only factor leading to increased aggregability of RBCs and modifications of membrane surface properties such as charge or protein distribution also lead to major changes in aggregability via changes in electrostatic or binding interactions. For instance it was shown that elevated concentrations of plasma sodium lead to impairment of the erythrocyte glycocalyx and alteration of the zeta-potential, resulting in higher aggregability \cite{McNally2023,Oberleithner2013} while on the other hand experiments with artificially rigidified cells suggest that a small fraction of these cells mixed with a healthy sample tends to reduce the aggegation index and mean aggregate size \cite{kuck2022impact}.

Even in healthy conditions, RBC properties are significantly heterogeneous and evolve in time. Produced in bone marrow, RBCs have a lifespan of about 120 days during which aging leads to increased density, stiffness and other alterations \cite{Waugh1992Mar,bosch1994,franco2012measurement,franco2013} before they are eliminated by phagocytosis by macrophages in the spleen.
As blood flows through the circulatory system, the stresses experienced by blood components vary as they navigate the complex geometries and varying diameters of the circulatory system \cite{caro2012mechanics,nichols2005mcdonald}. These flow events can cause RBCs to elongate, alter membrane protein distribution, and potentially lead to membrane fatigue, impacting cell functionality and lifespan \cite{lipowsky2005microvascular,pivkin2008accurate}.
As the cells age, they undergo several physical and chemical changes including loss of constituents like - water, (2,3)- bisphosphosphoglyceric acid, ATP, proteins, Hb, sialoglycoproteinssialic acids (SA), resulting structural changes to the cell membrane \cite{huang2011human, lutz1983senescent,lutz1987naturally}.

\rev{Many studies have used Percoll or Optiprep (iodixanol) density gradients to separate RBCs by density, aiming at correlating density changes, cell properties and aging. Percoll has been used for more than 40 years for density based separation of different kinds of cells without evidence of cell alteration. Recent works also specifically tested the deformability of RBCs comparing the characteristics of native cells and those that underwent Percoll processing and concluded that exposure to Percoll does not lead to alterations in RBC deformability \cite{Barshtein2024}, suggesting that it is a generally innocuous cell separation technique with few specific exceptions \cite{Wakefield1982}.  These separation techniques show that RBC density distributions are relatively wide, ranging from 1.08 to 1.12 g/mL. 
Recent works have shown that RBC density significantly correlates with their dynamics, demonstrating a significant impact on RBC shape in flow \cite{nouaman2023effect}, as well as migration and axial dispersion in capillaries \cite{losserand2019migration, losserand2023axial}. }

While there is no consensus on a strict equivalence between red cell density and age and studies show that there is a wide distribution of densities among cells of the same age, it is still accepted that on average density increases with cell age \cite{morrison1983does}. \rev{For instance, Franco et al. \cite{franco2013} have shown that the density distribution of biotin-labeled RBCs determined at successive moments in time by density fractionation using Optiprep (Iodixanol) density gradients was significantly shifted to higher densities over the course of 120 days. A robust feature is that cell density is related to hemoglobin concentration, which tends to increase as cells age due to dehydration \cite{mohandas86scattering, Vaysse88}}. 
In addition, density increase was found to be correlated to loss of membrane surface area. Both phenomena lead to a decrease of static deformability (deformability index) and dynamic deformability (increase of relaxation time). Besides, density increase is also observed in RBC disorders associated with decreased cell deformability \cite{mohandas88densdistrib}.

\rev{A few studies revealed that RBC aging leads to increased aggregation \cite{rampling2004influence}. 
In one of the earliest study,  Nordt \cite{1983_Nordt} showed that in autologous plasma, denser cells aggregated more than twice as much as younger, less dense cells, with intermediate-density cells displaying moderate aggregation. These results were extended to aggregation of density separated cells in 70 kDa dextran solutions by Nash et al. \cite{Nash1987}. However these early results are relatively qualitative in nature and do not provide a detailed assessment of aggregation characteristics other than variations in a global aggregation index \cite{chien1973ultrastructural}}.
Various techniques have been proposed to quantify aggregation and characterize the aggregability of erythrocytes at the suspension scale. These include the classic erythrocyte sedimentation rate (ESR) test \cite{bedell1985}, \rev{techniques based on light scattering such as} ektacytometry \cite{dacosta2016} or LORCA \cite{hardeman1994laser}, and ultrasound scattering \cite{franceschini2010ultrasound}, all of which are applied to dense suspensions. \rev{A limitation of these techniques is that they provide average measurements over the whole sample but fail to offer insight into the distribution of properties within a sample. Furthermore, these methods typically operate in simple shear flow, which does not fully represent conditions in the microcirculation.} At the cellular scale, force measurements or controlled force application methods, such as micropipette aspiration techniques \cite{tozeren1989theoretical, buxbaum1982quantitation}, AFM \cite{steffen2013quantification}, and more recently optical tweezers (OT) \cite{lee2017assessment, Ermolinskiy2020Oct}, allow direct investigation of cell-cell interactions.  Recently, Ermolinskiy et al. \cite{Ermolinskiy2020Oct} showed that the aggregation force measured by OT increases with the density of erythrocytes, both in autologous plasma and in 50 mg/ml Dextran 70 kDa solutions, \rev{confirming} an increase of RBC aggregability with cell age. \rev{However, this type of time-consuming aggregation force measurement only allows to measure a limited number of aggregates and does not account for hydrodynamic stresses that RBCs experience in vivo, limiting the applicability of their results. Finally, flow cytometry has been proposed as a way to measure RBC aggregation \cite{Zhao_cytometry_aggregation2017} but also operates over a limited range of flow stresses that may not adequately represent in vivo flow conditions. Therefore, there remains a notable absence of quantitative techniques capable of analyzing the aging-induced variations in these interactions across a large number of aggregates in realistic flow conditions.}

Recently, a new microfluidic technique combined with machine learning analysis was proposed to investigate the dissociation of small RBC aggregates in confined elongational flows relevant to the hydrodynamic stresses experienced in microcirculatory networks \cite{puthumana2024dissociation}. This high-throughput technique provides detailed information about dissociation statistics as a function of strain rate for aggregates of 2 to 4 cells over large samples, complementing other techniques.

In this work, we use this technique to quantify the variations in dissociation probabilities of small aggregates of healthy erythrocytes in extensional flow as a function of cell density after separating them using Percoll density gradients. \rev{We provide a detailed and quantitative description of RBC aggregate strength as a function of RBC density, confirming
that aggregates formed from the denser fractions of red blood cells are significantly more robust than the average cell population and require significantly higher flow stresses for dissociation, and providing the dispersion of these values in a given sample}. These results \rev{go beyond a simple correlation between RBC age and aggregation and add substantial value to the understanding of RBC aggregation dynamics in microcirculatory conditions. In addition, they demonstrate the relevance of the technique for improved evaluation of RBC aggregability in health and disease and provide valuable experimental data for the validation of numerical models.}

 \section{Methods}
 
\subsection{Blood sample collection and preparation} 

\begin{figure}[t!]
     \centering
 \includegraphics[width = 0.8\linewidth, ]{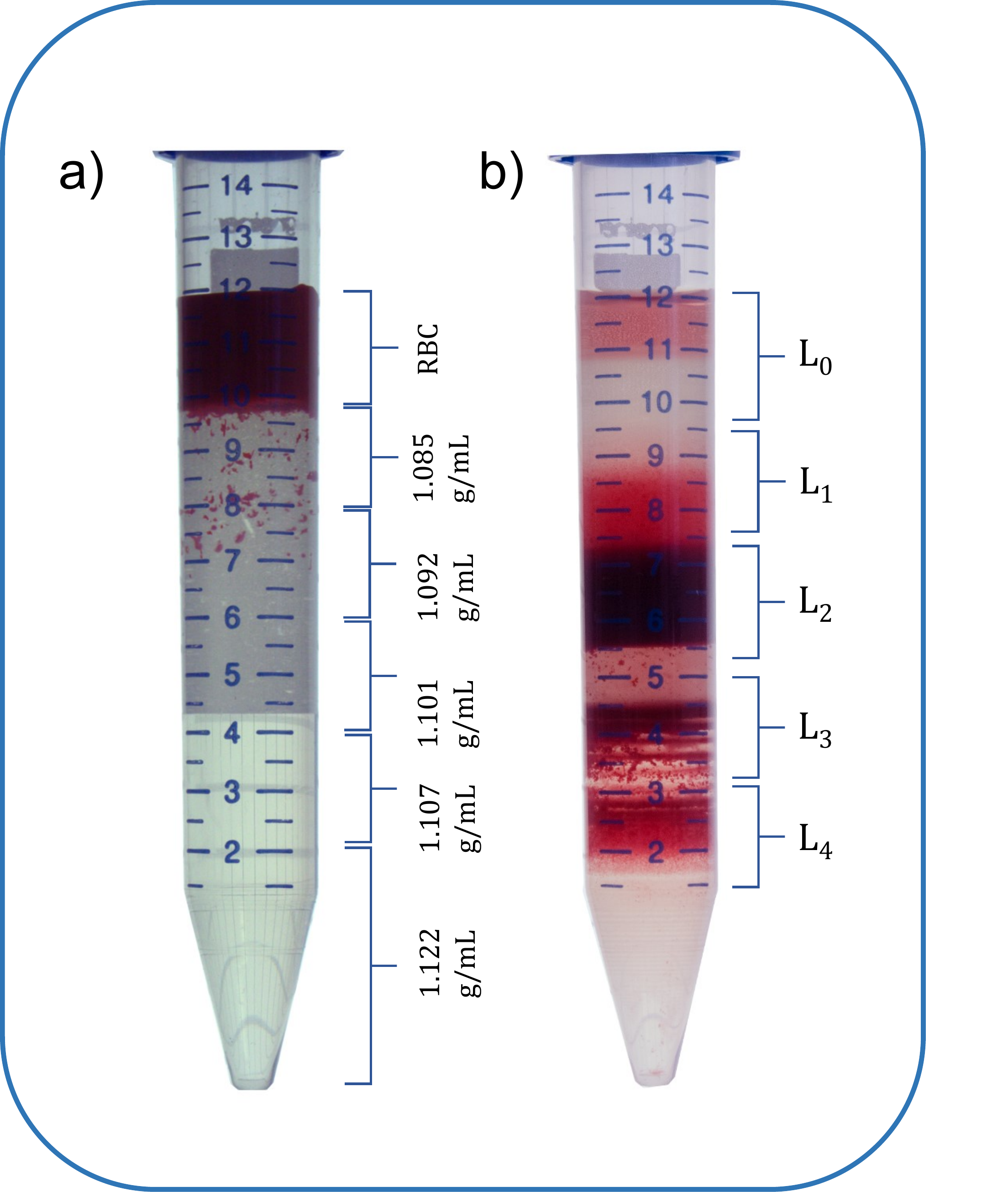}
   \caption{RBC density fractionation using percoll solution gradient. a) 2 mL of 60\% Hematocrit along with PBS is carefully placed above the $5 \times 2$~mL Percoll density gradient solution. The densities of each layer is shown in the figure. b) Separated RBC layers, after centrifugation for 30 minutes at 4000 $\times$ g. }
     \label{fig:density_separation}
 \end{figure}

Blood samples were collected from three healthy male volunteers with informed consent through venipuncture, using tubes containing EDTA (1.6 mg/mL, SARSTEDT, N\"{u}mbrecht, Germany) as anticoagulant. 
Plasma and buffy coat were then separated from red blood cells by centrifugation (3000g, 9 minutes) prior to washing 3 times (3000g, 3 minutes) using PBS (Phosphate Buffered Saline, Gibco, pH 7.4, Fisher Scientific, Germany).

The separation of RBCs according to cell age was conducted by utilizing the density variations that arise during the ageing process \cite{Waugh1992Mar}. To obtain density fractions of RBCs from a whole blood sample, we employed the Percoll density gradient centrifugation method outlined by Ermolinskiy et al. \cite{Ermolinskiy2020Oct}. The Percoll solution (Cytiva 17-0891-01, Sigma-Aldrich, Taufkirchen, Germany) consists of Percoll, distilled water, and a 1.5M NaCl solution in different proportions \rev{adjusted to reach isotonicity with a constant measured osmolarity of $300 \pm 5$ mOsm (detailed compositions are provided as Supplemental Material). This yields} the five different density levels for the experiment. Five solutions, with densities of 1.085 g/mL, 1.092 g/mL, 1.101 g/mL, 1.107 g/mL, and 1.122 g/mL, were prepared and carefully placed one over the other in a 15 mL centrifugation tube (highest density solution at the bottom, lowest at the top, with each layer of volume of 2 mL). Finally, a 2 mL suspension of washed red blood cells, diluted to a hematocrit of 60\% in PBS, was carefully layered onto the Percoll gradient solution. The sample was then centrifuged at 4000 g for 30 minutes at a controlled temperature of 4°C. Centrifugation results in cells being fractionated and forming bands labelled $L_0-L_4$ in the transition zones between Percoll layers as shown in Fig. \ref{fig:density_separation}, with $L_2$ being the most concentrated layer i.e. corresponding to the peak of RBC density distribution in blood samples \cite{mohandas88densdistrib,Ermolinskiy2020Oct,nouaman2023effect}. These different layers were then pipetted, washed, centrifuged and resuspended in a solution of 3 g/dL Dextran 70kDa to promote RBC aggregation with an interaction energy that is comparable to fibrinogen concentrations in the 4-5 g/L range \cite{neu2002,steffen2013quantification,brust14} before using them separately in experiments. Due to the very low cell concentrations in layers $L_0$ and $L_1$, only layers $L_2$, $L_3$ and $L_4$ were used in experiments. 
\rev{While establishing a direct relationship between RBC density and age is challenging due to the wide dispersion of densities among cells of the same age, as shown in previous studies, it is still possible to estimate age ranges for each density layer. This can be achieved using data from Franco et al. \cite{franco2013} in combination with cell lifespan models, such as those proposed in \cite{shrestha2016models}, applying Bayes’ theorem to generate histograms of age versus density. However, due to the discrete nature of the original datasets, these estimates are not highly precise and should be interpreted with caution. Acknowledging this limitation, we estimate that the studied layers approximately correspond to the interquartile age ranges indicated in Table \ref{tab:layers}.}

\begin{table}[h]
    \centering
    \begin{tabular}{l|c|c}
        \hline
        Layer & Density Range (g/mL) & Age Range (days) \cite{franco2013,shrestha2016models} \\
        \hline
        $L_2$ (light) & 1.092--1.101 & $40 \pm 30$ \\
        $L_3$ (Intermediate) & 1.101--1.107 & $53 \pm 28$ \\
        $L_4$ (dense) & 1.107--1.122 & $58 \pm 28$ \\
        \hline
    \end{tabular}
    \caption{Density layers studied in the experiment and corresponding density ranges and estimated age ranges.}
    \label{tab:layers}
\end{table}

\rev{Note that during the density separation process, RBCs were exposed to Percoll for no more than two hours. While Percoll has been shown to promote aggregation when cells are suspended in this medium \cite{maurer2022continuous}, it is completely removed after washing individual layers in PBS, with no effect on aggregability in the experiment. To rule out any potential impact of Percoll on aggregation and RBC mechanical properties, we independently verified that RBCs exposed to Percoll showed no measurable difference in aggregation compared to RBCs from the same sample that had not been in contact with Percoll (see Supplemental Material). Layers were studied in a random order and kept at 4\textdegree C while waiting to be studied in the microfluidic setup for a time course of 2-4 hours, making up a total delay between collection and study of no more than 6 hours.}


\subsection{Microfluidics}

\begin{figure}[t!]
    \centering
   \includegraphics[width = 0.95\linewidth, ]
   {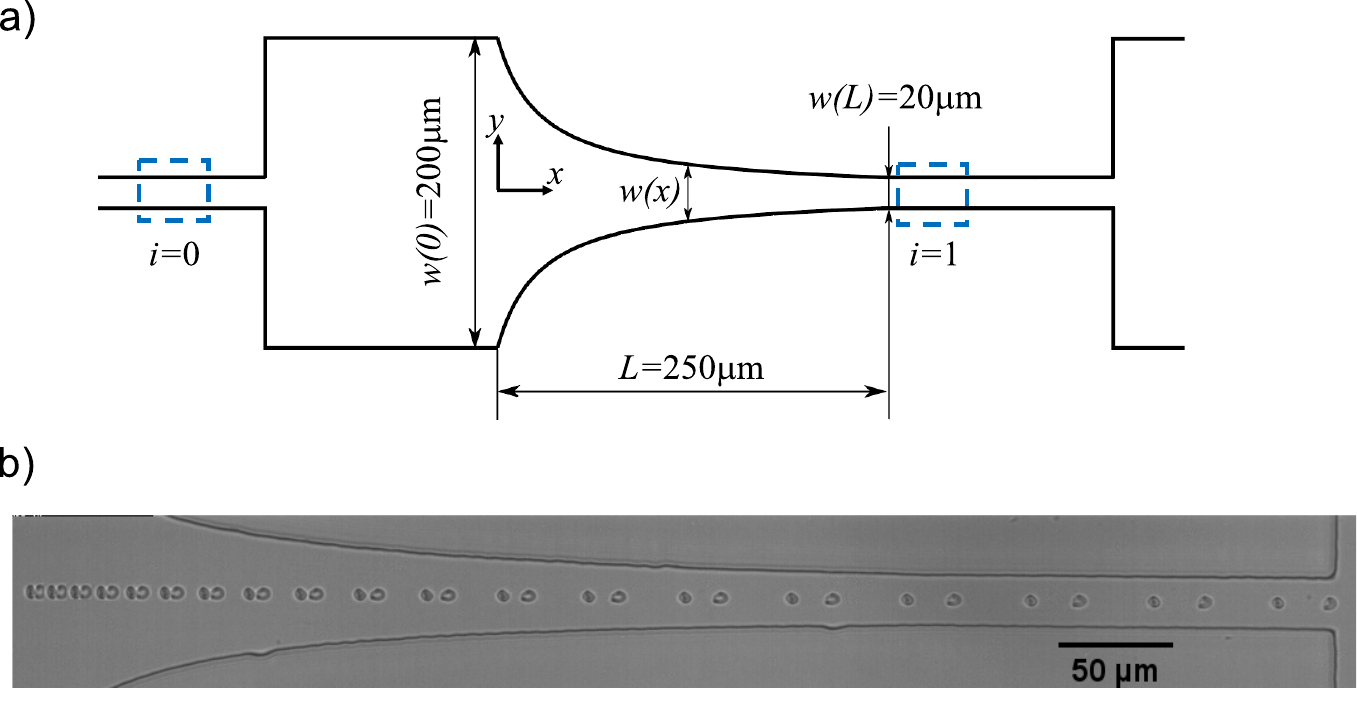}
    \caption{Hyperbolic flow geometry producing extensional flow: (a) Design of the central part of the microfluidic channel and notations. Blue areas are image processing ROIs;   (b) 
    Example temporal sequence of a 2-cell aggregate breaking into $1+1$ cells under extensional flow. Time-step: 1.1  ms; Extension rate $\dot{\varepsilon}=$ 118 s$^{-1}$.
    }
    \label{fig:constriction}
\end{figure}

The core component of the microfluidics setup, whose principle was described in detail in a previous study and was demonstrated to be a reliable tool to characterize RBC disaggregation \cite{puthumana2024dissociation}, is a hyperbolic constriction geometry (Fig. \ref{fig:constriction} a). This design has been effectively implemented in prior studies \cite{bento2018deformation,faustino2019microfluidic} to induce an extensional flow thanks to a channel width that decreases inversely with the coordinate $x$ in the flow direction ($w(x)\sim 1/x$) and a constant channel thickness $h$. Mass conservation implies that the cross-sectional average of the axial velocity $<u_x(x)>$ increases linearly along $x$ i.e. the average extension rate $d<u_x(x)>/dx$ is constant.
Particles flowing along the central line therefore experience an extensional strain rate $\dot{\varepsilon_f} = {d v_x}/{dx}$ where $v_x=u_x(x,0,0)$. A series of such constrictions with constant thickness $h=20$~$\mu$m, in the $z$ direction, inlet width($w_0= 200$~$\mu$m) and outlet width ($w_L = 20$~$\mu$m) was implemented, with decreasing lengths, $L$ (1000~$\mu$m, 500~$\mu$m, 250~$\mu$m, 125 ~$\mu$m), inducing increasing extension rates $\dot{\varepsilon_f}$ for a given flow rate. The measurements of aggregate dissociation statistics presented here were conducted in the hyperbolic channel with $L = 250$ $\mu$ m as in \cite{puthumana2024dissociation}.

The entrance section of the microfluidic chip is a 3-inlet flow-focusing device where the concentrated RBC suspension is focused in the $y$-direction between two perpendicular inlets of the suspending medium. It is followed by a $\sim$ 5~cm long channel, which allows RBC aggregates to relax and focus around the center plane in the $z$-direction thanks to migration forces \cite{losserand2019migration,grandchamp13,podgorski11}. The flow is driven by applying pressure at 3 inlets of the microfluidics chip, using an Elveflow OB1 pressure controller.

To facilitate image acquisition, each constriction is followed by a straight channel (cross section $w(L)\times h$, length 150$\mu$m). As discussed in our previous work demonstrating the experimental principle \cite{puthumana2024dissociation}, RBC aggregates are insensitive to shear in this confined configuration and the evolution of aggregate size statistics can be attributed solely to extensional stresses created by the constriction.

\subsection{Microscopy and Image processing}

RBC cluster images were acquired using an inverted microscope (Eclipse TE2000-S, Nikon) equipped with a x20 objective lens and a high-speed camera (Fastec HiSpec 2G, FASTEC Imaging), providing a 2.05 pixels/$\mu$m resolution and a frame rates ranging 250-3000 fps. A typical aggregate dissociation sequence is shown in Fig. \ref{fig:constriction}(b). The actual extension rate experienced by aggregates $\dot{\varepsilon}=dv_p/dx$
where $v_p$ is the particle velocity was evaluated for each value of the inlet pressures by measuring sequential positions $x(t)$ of a few particles (cells or aggregates) along the hyperbolic constriction and fitting it with an exponential function $x(t)=a+b \exp(\dot{\varepsilon} t)$.

The cluster size distributions before and after the constriction were evaluated from image sets containing between 1000 and 10000 objects (cells and clusters) for each experiment sequence, using a combination of classical image processing and a custom convolutional neural network (CNN) algorithm described in \cite{puthumana2024dissociation} and based on the TensorFlow framework \cite{abadi2016tensorflow}.
After training the CNN with a carefully assembled set of images, achieving an impressive accuracy of approximately 95\%, the algorithm classifies flowing objects into single cells, doublets, triplets, quadruplets, and larger objects, with their respective relative populations denoted as $S_i$, $D_i$, $T_i$, $Q_i$, and $N_i$ (where $i=0$ before and $i=1$ after the constriction as defined in Fig. \ref{fig:constriction}a). These populations are normalized by the total number of detected RBCs (aggregated or not), ensuring that $S_i+2D_i+3T_i+4Q_i=1$. Aggregates of five cells or more are very rare and are not included in this normalization.

Assuming that aggregates of 5 cells or more are negligible ($N_i \ll 1$), mass conservation between the initial size distribution before the constriction $(S_0, D_0, T_0, Q_0)$ and after the constriction $(S_1, D_1, T_1, Q_1)$ allows to compute dissociation probabilities $P_d$ (for doublets),  $P_t$ (triplet breaking into a doublet and a single cell), $P_{q1}$ and $P_{q2}$ (quadruplet breaking into respectively two doublets or a single cell and a triplet), with $P_q=P_{q1}+P_{q2}$, with the additional assumption that $P_{q1}=2 P_{q2}$ (assuming that all cell-cell bonds in a quadruplet are equivalent) as done previously \cite{puthumana2024dissociation}.




\begin{figure*}[ht!]
 \resizebox{1\textwidth}{!}{
   \includegraphics{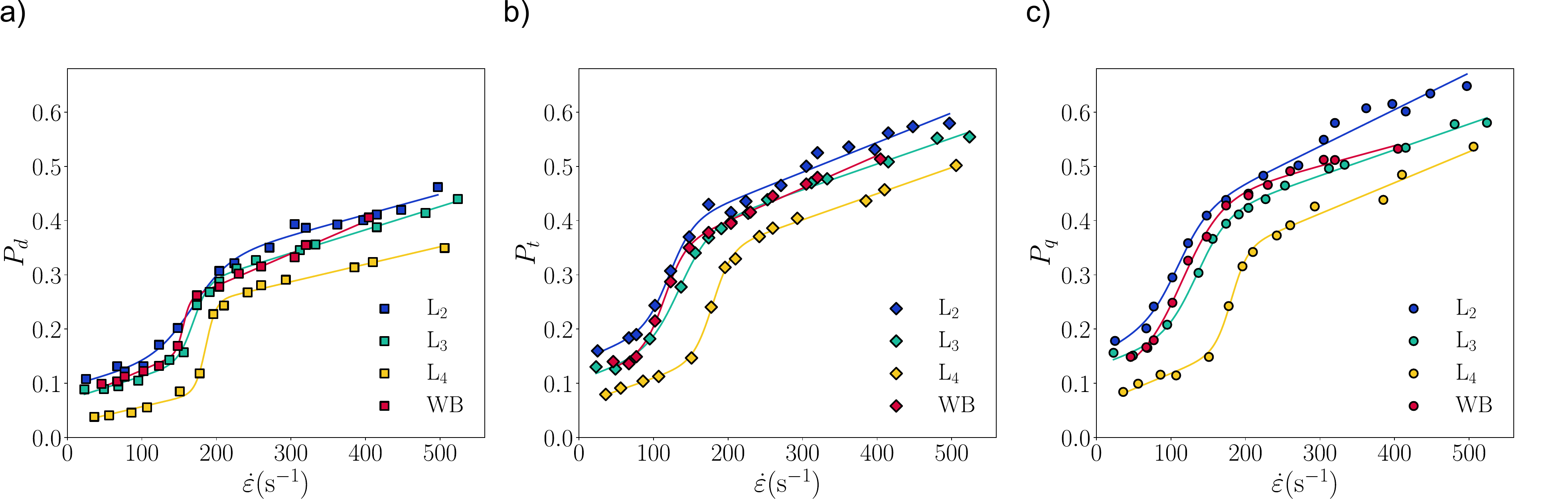}}
  \caption{ Dissociation probabilities of aggregates of 2, 3 and 4 cells (doublets (a), triplets (b) and quadruplets (c)) for RBCs from layers $L_2$ to $L_4$ obtained by density fractionation and from whole blood (WB) samples (data from \cite{puthumana2024dissociation}). Continuous lines represent fitting curves using equation \ref{eqnfit}. All  series represent data from 3 different donors. 
  \label{fig:breakingstats_saar}}
\end{figure*}

\section{Results and Discussion}

\begin{figure*}[ht!]

 \resizebox{0.7\textwidth}{!}{
   \includegraphics{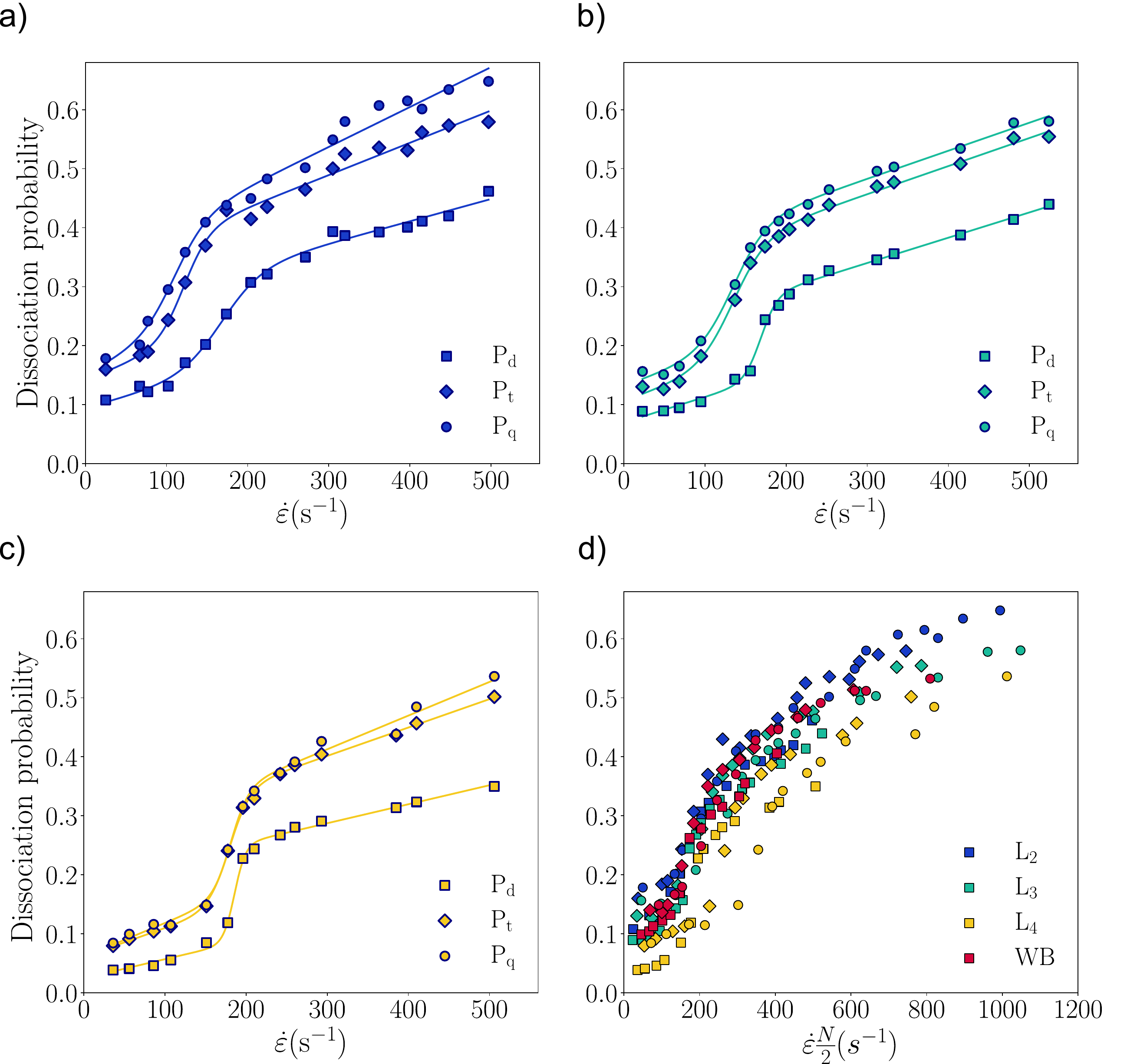}}
  \caption{Dissociation probabilities of aggregates of 2, 3 and 4 cells ( $\blacksquare$ : doublets,$\Diamondblack$: triplets, $\medbullet$: quadruplets) of RBCs from  layers (a) $L_2$ (b) $L_3$ (c) $L_4$ obtained by density fractionation vs. extension rate, with fitting curves using equation \ref{eqnfit}. (d): Data from all layers vs. $\dot{\varepsilon}N/2$ as suggested by Eq. \ref{eqn:crit_extension} and compared to data for whole blood (from \cite{puthumana2024dissociation}).  \label{fig:breakingstats_layerwise}}
\end{figure*}

The results are summarized in Fig. \ref{fig:breakingstats_saar} in which the dissociation probabilities $P_d$, $P_t$, $P_q$ of aggregates of cells from density layers $L_2$ to $L_4$  and from whole blood ($WB$) are plotted as a function of extension rate $\dot{\varepsilon}$, with a 3g/dL Dextran buffer as suspending medium.

As previously reported \cite{puthumana2024dissociation}, all disaggregation probabilities follow a characteristic sigmoid curve when increasing the extension rate. \rev{An assessment of inter-donor variability shows that differences between healthy samples are minimal, with all data points aligning on the same master curve, regardless of donor or the order in which different layers were studied. This also confirms that fluctuations in the delay between sample collection and analysis had no significant effect within the time frame of our experiments (see Supplemental Material).} A strong jump observed around the extension rates of $80-200$~s$^{-1}$ allows to define a critical extension rate $\dot{\varepsilon}_c$ corresponding to the maximum rate of change in dissociation probabilities. To do so, experimental data in Fig. \ref{fig:breakingstats_saar} is fitted with a sigmoid curve defined as:
\begin{equation}
P(\dot{\varepsilon})= A + B \tanh \left(C(\dot{\varepsilon}-\dot{\varepsilon}_c)\right) + D \dot{\varepsilon} ~,
\label{eqnfit}
\end{equation}
where $A$, $B$, $C$, $D$ and $\dot{\varepsilon}_c$ are fitting parameters. The determination coefficient $R^2$ for all data series in Fig. \ref{fig:breakingstats_saar} is better than $0.99$, allowing for a reliable determination of $\dot{\varepsilon}_c$.

In \cite{puthumana2024dissociation}, an analysis of the forces acting on cells within a cluster, performed in a reference frame centered on the cluster’s center of mass, concluded that the dissociation dynamics are primarily governed by the balance between the work of viscous drag forces on the cells and the interaction energy between them. This led to the proposal of a scaling law for the critical extension rate required to dissociate a two-cell aggregate (doublet):

\begin{equation}
\dot{\varepsilon}_c \simeq \frac{\epsilon_{ad}}{12 \eta \delta l} ~,
\label{eqn:crit_extension}
\end{equation}
which can be generalized for a (linear) aggregate of $N$ cells:

\begin{equation}
\dot{\varepsilon}_c(N) \simeq \frac{\epsilon_{ad}}{12 (N/2) \eta \delta l} ~,
\label{eqn:crit_extension}
\end{equation}
where $\epsilon_{ad}$ is the interaction energy per unit area, $\eta$ is the suspending medium's viscosity and $\delta l$ is the typical displacement required to separate cells. Note that for a doublet ($N=2$) with $\epsilon_{ad} = 5~\mu\text{J/m}^2$ \cite{steffen2013quantification}, \(\eta = 2~\text{mPa.s}\) (viscosity of 3g/dL dextran solution), and \(\delta l \approx 1~\mu\text{m}\) (displacement to separate cells), equation. \ref{eqn:crit_extension} gives a critical extension rate of about 200 s\(^{-1}\), which is the order of magnitude observed in experimental measurements (Fig. \ref{fig:breakingstats_saar}).

As suggested by the scaling $\dot{\varepsilon}_c(N) \sim 1/N$, larger aggregates tend to dissociate at lower extension rates compared to smaller aggregates as larger objects experience higher drag forces in the extensional field, as seen in Fig. \ref{fig:breakingstats_layerwise}(a-c) where data for each layer is represented separately. 
This is also reflected in Fig. \ref{fig:breakingstats_layerwise}(d) where dissociation probabilities are plotted as a function of $\eta \dot{\varepsilon} N/2 $ as suggested by Eq. \ref{eqn:crit_extension}: data for triplets ($P_t$) and quadruplets ($P_q$) rather nicely collapse with the data for doublets ($P_d$) from $L_2$, $L_3$ and $WB$. The collapse is not as good for $L_4$ (cells with the highest density - corresponding to older cells) for which   $\dot{\varepsilon}$ only weakly depends on $N$ as visible in Fig. \ref{fig:breakingstats_layerwise}(c) where the lateral shift is smaller and dissociation probabilities for triplets and quadruplets are almost identical. We shall discuss this below.


\begin{figure}[ht!]
 \resizebox{0.95\columnwidth}{!}{
   \includegraphics{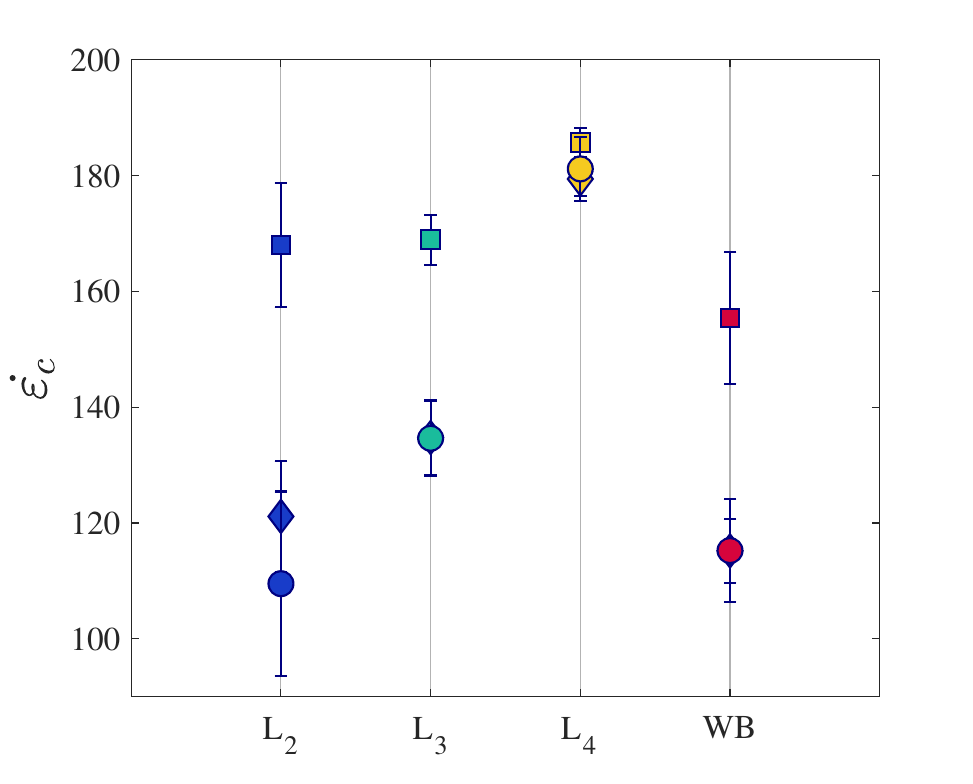}}
  \caption{ Critical extension rates for aggregates of different layers ( $\blacksquare$ : doublets,$\vardiamondsuit$: triplets, $\medbullet$: quadruplets) estimated by fitting Eq. \ref{eqnfit} to  the data represented in Fig. \ref{fig:breakingstats_layerwise}. Error bars represent 95\% confidence intervals defined as $1.96 \sigma$ where $\sigma$ is the standard error of the fitting parameter $\dot{\varepsilon}$.
  \label{fig:crit_extension}}
\end{figure}

Comparing the data from different density layers in Figs. \ref{fig:breakingstats_saar} and \ref{fig:breakingstats_layerwise} shows that RBCs from denser layers tend to form more robust aggregates that require higher extensional stresses to dissociate. When increasing $\dot{\varepsilon}$, cells from $L_2$ are the first to dissociate, followed by $L_3$, and then by $L_4$.  The experimental critical extension rate $\dot{\varepsilon}_c$ defined as the inflection point in dissociation curves of Fig. \ref{fig:breakingstats_saar} - \ref{fig:breakingstats_layerwise} is shown in Fig. \ref{fig:crit_extension}. On average, aggregates from $L_4$ require extensional stresses that are approximately twice as high as the value needed for $L_2$ to reach a similar dissociation probability (e.g. for quadruplets) with a significant increase (up to 65\%) of $\dot{\varepsilon}_c$. Note that the data for $WB$ is roughly between the data for $L_2$ and $L_3$ in Figs  \ref{fig:breakingstats_saar} and \ref{fig:breakingstats_layerwise}(d), which is consistent with the fact that $L_2$ corresponds to the peak of the RBC density distribution as seen in Fig. \ref{fig:density_separation} and previous works \cite{mohandas88densdistrib,Ermolinskiy2020Oct,nouaman2023effect}. 

The very strong shift in extension rate observed for $L_4$ compared to a more modest one between $L_2$ and $L_3$ is reminiscent of the variation of deformability reported by Nouaman et al. \cite{nouaman2023effect} who measured deformability indexes (DI) and DI variations with strain rate that are drastically reduced for $L_4$ using the same density fractionation, or with earlier ektacytometry measurements that also show more pronounced changes for the densest fractions of RBCs \cite{mohandas1989cell}. In comparison, optical tweezer measurements  \cite{Ermolinskiy2020Oct} revealed a significant and gradual increase of about 20\% of the aggregation force between $L_2$ and $L_4$ in autologous plasma and about 10\% in a 5g/dL Dextran 70 kDa solution. Although the suspending media are different, the amplitude of the measured effect are comparable with the evolution of $\dot{\varepsilon}_c$ for doublets in Fig. \ref{fig:crit_extension}: the critical extension rate for doublets  is about 10\% higher in $L_4$ compared to $L_2$. In parallel, the value of $P_d$ is also significantly lower in $L_4$ at any value of $\dot{\varepsilon}$ (see Fig. \ref{fig:breakingstats_saar}), reflecting stronger aggregability.
The aggregate sizes of $L_4$ cells have been shown to be smaller compared to other layers in  cytometry measurements of Percoll separated samples \cite{huang2011human}. Following the analysis leading to Equation \ref{eqn:crit_extension}, aggregates of older cells may therefore require a higher extension rate to experience the same total drag force as slightly bigger young cell aggregates.


The evolution with cell density is however much more contrasted with aggregates of 3 and 4 cells that are much more robust in $L_4$. This is also consistent with the fact that in this density layer, aggregates of different sizes (doublets, triplets, quadruplets) dissociate at about the same extension rate and do not obey the scaling with $N$ proposed in equation.\ref{eqn:crit_extension}. This can be related to morphological changes that can be observed and have been documented for denser and older cells:
\begin{itemize}
\item cell aging and density increase is associated to dehydration, membrane loss and other alterations that can lead to smaller, less deformable and possibly more spherical cells. A higher occurence of echinocytes in older cells has also been reported \cite{huang2011human}. These alterations at the cell level have consequences on aggregate morphology.
\item for the same reason as mentioned above, the higher compacity or smaller volume of aggregates of denser cells means that the scaling with $N$ in Eq. \ref{eqn:crit_extension} is less applicable, as it is based on the assumption that aggregates are linear rouleaux.
 \item differences in membrane properties such as surface charge are also likely to contribute \cite{hadengue1998erythrocyte} in the variations observed in between younger (lighter) and older (denser) cell aggregates.
\end{itemize}


\section{Conclusions}


We investigated the dissociation of red blood cell aggregates under extension, focusing on the effects of cell aging, using cell density as an indicator of cell age. A significant variation in the dissociation of aggregates with cell density was observed, both in terms of critical hydrodynamic stress and dissociation probability. This finding is consistent with previous studies on the differences in aggregation strengths between cells of varying ages and confirms that RBC aggregability increases with cell density and age.

In contrast to other aggregability measurement methods, the microfluidic technique we used here provides hydrodynamic conditions that are relevant to flow in the microcirculation, particularly in the bifurcations of capillary networks. Our findings therefore contribute to the understanding of aggregation dynamics in microcirculation and the consequences of cell aging. While it is well known that cell aging in vivo leads to decreased cell deformability, which is the mechanism by which older cells are filtered and removed from the blood in the spleen \cite{Moreau2023}, the increased RBC aggregability and aggregate robustness that we quantified in this work are likely to have significant consequences at the microcirculatory level (increased resistance to flow, increased clogging) that may be more severe than the simple decrease in deformability for single cells. This further underscores the physiological importance of removing aging cells from circulation.  

Altered properties through which aggregability and robustness of aggregates may be increased for denser and older cells were mentioned, namely RBC shape, volume and surface changes, decreased deformability and altered membrane surface properties. Although these effects are not dissociable in our study, we suggest that the same technique can be used to study the influence of each parameter separately, for instance after modifying RBC deformability in a controlled way using diamide or glutaraldehyde, or altering RBC volume by suspending them in hyper or hypo-osmotic media. 


\begin{acknowledgments}

The authors thank M. Van Melle-Gateau (LIPhy, CNRS-UGA) for technical assistance (microfabrication), T. John, M. Nouaman, A. Darras and S. Reckenwald  for experimental advice; M. P M and T.P. thank G. Ghigliotti, E. Franceschini and G. Coupier for discussions and suggestions. This work was supported by CNES (Centre National d'Etudes Spatiales, DAR ID 8106 "Rh\'eologie sanguine"), and ANR project ANR-15-IDEX-02 (IDEX Universit\'e Grenoble Alpes, "Aide \`a la mobilit\'e internationale des doctorants"). 
\end{acknowledgments}

\bibliographystyle{unsrt}

\bibliography{Aging_Disaggregation}

\end{document}